\RequirePackage{fix-cm}
\documentclass[smallextended]{svjour3}       
\pdfoutput=1
\usepackage{mathptmx}
\usepackage{graphicx}
\usepackage{natbib}
\usepackage{amssymb, amscd, amsfonts, amsmath}
\usepackage{srcltx}
\usepackage{bm}
\usepackage[T1]{fontenc}
\usepackage[utf8]{inputenc}
\usepackage[english]{babel}
\usepackage[pdftex,dvipsnames]{xcolor}
\usepackage{xargs}
\usepackage[colorinlistoftodos,prependcaption,textsize=scriptsize]{todonotes}
\usepackage{siunitx}
\usepackage{setspace}
\newcommand {\RR } {{\mathbb R}}
\newcommand {\bx } {{\boldsymbol x}}
\newcommand {\by } {{\boldsymbol y}}

\newcommand {\bu } {{\boldsymbol u}}
\newcommand {\bg } {{\boldsymbol g}}
\newcommand {\bv } {{\boldsymbol v}}
\newcommand {\bs } {{\boldsymbol s}}
\newcommand {\bw } {{\boldsymbol w}}
\newcommand {\bh } {{\boldsymbol h}}
\newcommand {\be } {{\boldsymbol e}}

\newcommand {\bz } {{\boldsymbol z}}
\newcommand {\bZ } {{\boldsymbol Z}}
\newcommand {\bC } {{\boldsymbol C}}
\newcommand {\bW } {{\boldsymbol W}}
\newcommand {\bA } {{\boldsymbol A}}
\newcommand {\bq } {{\boldsymbol q}}
\newcommand {\dist } {d}

\def\refe#1{(\ref{#1})}

\definecolor{darkgreen}{rgb}{0.05,0.05,0.5}
\definecolor{darkred}{rgb}{0.5,0.05,0.05}
\definecolor{darkblue}{rgb}{0.05,0.05,0.5}

\newcommand {\pd}[2] {\frac{\partial #1}{\partial #2}}
\newcommand {\pddm}[2] {\frac{\partial^2 #1}{\partial {#2}}}
\newcommand {\pddl}[2] {{#1}_{#2}}
\newcommand {\der}[2] {d_{#2}#1}
\newcommand {\fd}[3] {\mathcal{D}#1(#2;#3)}
\newcommand {\sd}[3] {\mathcal{D}^2#1(#2;#3)}
\newcommand {\jac}[2] {J_{#2}(#1)}
\newcommand {\jact}[2] {J^t_{#2}(#1)}

\newcommand{\var}[1]{{\bh}} 

\newcommand {\dint}[1] {\,\mathrm{d}#1}
\newcommand {\dt}[1] {\dint{t}}

\newcommand {\scal}[2]{\left(#1,#2\right)}

\newcommand {\dual}[2]{\left<#1,#2\right>}
\newcommand {\dualk}[3]{\left<#1,#2\right>_{#3}}

\newcommand {\norma}[1]{\left\|#1\right\|}

\newcommand{\mle}{\hat{\pars}^{MLE}}

\newcommand{\nom}{N} 
\newcommand{\meas}{\by} 
\newcommand{\mf}{\bg} 
\newcommand{\rese}{\boldsymbol{\varepsilon}} 
\newcommand{\pars}{\boldsymbol\phi} 
\newcommand{\dimo}{n}
\newcommand{\dimpars}{p} 
\newcommand{\recov}{\Sigma} 
\newcommand{\llf}{l} 

\newcommand{\mv}{{\bu}}
\newcommand{\proj}{{\mathcal P}}
\newcommand{\dimmv}{m}
\newcommand{\mrhs}{{\boldsymbol f}}
\newcommand{\sv}{\bs}
\newcommand{\dv}{\bv}
\newcommand{\ssv}{{\boldsymbol\varsigma}}

\newcommand{\tni}{T_{NI}}
\newcommand{\tl}{T_{L}}
\newcommand{\ta}{T_{A}}
\newcommand{\vi}{V_{I}}
\newcommand{\vni}{V_{NI}}

\newcommand{\enrti}{\eta_{NRTI}}
\newcommand{\epi}{\eta_{PI}}
\newcommand{\muni}{\mu_{NI}}
\newcommand{\mul}{\mu_{L}}
\newcommand{\al}{\alpha_{L}}
\newcommand{\mua}{\mu_{A}}
\newcommand{\muv}{\mu_{V}}

\newcommand{\diffmem}{\textit{Diff}MEM}

\authorrunning{Valdemar Melicher et al.}

\smartqed  

\begin{document}

\def\spacingset#1{\renewcommand{\baselinestretch}%
{#1}\small\normalsize} \spacingset{1}

\title{\bf Fast derivatives of likelihood functionals for ODE based models using adjoint-state method\thanks{The work of the first two authors was supported by IWT O\&O project 130406 -- ExaScience Life HPC.}}

\titlerunning{Fast derivatives of likelihood functionals for ODE based models using ASM}

\author{Valdemar Melicher$^1$ \and Tom Haber$^2$ \and Wim Vanroose$^1$}


\institute{
\\[-2mm]
Valdemar Melicher\\
\email{Valdemar.Melicher@UAntwerpen.be}\\[2mm]
$^1$ Department of Mathematics and Computer Science, University of Antwerp, Middelheimlaan 1, 2020 Antwerp, Belgium\\
$^2$ Expertise Centre for Digital Media, Hasselt University, Wetenschapspark 2, 3590 Diepenbeek, Belgium
}

\date{Received: date / Accepted: date}

\maketitle

\begin{abstract}
We consider time series data modeled by ordinary differential equations (ODEs), widespread models in physics, chemistry, biology and science in 
general. The sensitivity analysis of such dynamical systems usually requires calculation 
of various derivatives with respect to the model parameters.

We employ the \emph{adjoint state method} (ASM) for efficient computation of 
the first and the second derivatives of likelihood functionals constrained by ODEs with respect to the parameters of the underlying ODE model. Essentially, the gradient 
can be computed with a cost (measured by model evaluations) that is
independent of the number of the ODE model parameters and the Hessian
with a linear cost in the number of the parameters instead 
of the quadratic one. The sensitivity analysis becomes feasible even if the 
parametric space is high-dimensional.

The main contributions are derivation and rigorous analysis of the ASM in the statistical context, when the discrete data are coupled with the continuous ODE model. Further, we 
present a highly optimized implementation of the results and its benchmarks on a number of problems.

The results are directly applicable in (e.g.) maximum-likelihood estimation or Bayesian sampling of ODE based statistical models, allowing for faster, more
stable estimation of parameters of the underlying ODE model.
\end{abstract}

\noindent%



\section{Introduction}

We consider time series vector data $\meas_{i}\in \RR^n$ for $i=1,\dots, \nom$, where $\dimo$ is the dimension of the observation 
space and $\nom$ is the number of corresponding measurements times $t_i$ in the interval $I:=[0,T]$ with some positive final time $T > 0.$ In many scientific fields, the underlying structural model for such data is very often an initial-value problem of the following type:

\begin{equation}
\begin{aligned}
\der{\mv}{t} &= \mrhs(t, \mv, \pars),\quad t\in [0,T],\\ 
\mv(0) &= \mv_0(\pars),
\end{aligned} 
\label{eq:ivp}
\end{equation}
where $\mv_0$ is the initial condition, dependent only on the parameter vector
$\pars \in \RR^\dimpars$. In general non-linear r.h.s. $\mrhs$ of the governing equation represents the time derivative of the model variable $\mv(t).$ It depends on the current time $t$, the model parameters $\pars$ and the current values of $\mv \in \RR^\dimmv.$

The predictor $\hat\meas$ of the data $\meas$ is a result of integration of the dynamical system \refe{eq:ivp} and a possible subsequent post-processing, for example aggregation. This can be expressed in mathematical terms as $\hat\meas = \proj(\mv(t, \pars)) =: \mf(t, \pars),$ where $\proj: \RR^\dimmv \to \RR^\dimo$ is the post-processing operator relating the solution $\mv$ to data.

The main aim of this paper is to efficiently compute the first and the second derivatives of log-likelihood functionals of the following form
\begin{equation}
l(\pars) = -\sum_{i} \dist(\meas_i, \mf(t_i,\pars)),
\label{eq:generallikelihood}
\end{equation}
with respect to $\pars.$ Here $\dist: \RR^n\times\RR^n \to [0,\infty)$ is a sufficiently smooth distance function (metric) on $\RR^n$. Equation \ref{eq:generallikelihood} measures the fidelity between the model and the data.

\subsection{Motivation}
The most prominent example of log-likelihood functional \refe{eq:generallikelihood} is obtained for error model
\begin{equation}
\meas_{i} = \mf(t_i, \pars) + \rese_i, \qquad \rese_i \sim_{i.i.d.} \mathcal{N}(0, 
\recov), 
\label{eq:gaussianerror}
\end{equation} 
i.e. the residual errors $\rese_i$ are independent and 
identically distributed normal random variables with zero mean and residual covariance matrix $\recov \in \RR ^{\dimo\times \dimo}.$ Then 
\begin{equation}
\dist(\meas_i, \mf(t_i,\pars)) := \frac12(\meas_i-\mf(t_i, \pars))^t\recov^{-1}(\meas_i-\mf(t_i, \pars))
\label{eq:l2weightedMetric}
\end{equation}
and we are interested in the derivatives of
\begin{equation}
l(\pars) := \log p(\meas|\pars) \propto -\sum_{i} \dist(\meas_i, \mf(t_i,\pars)).
\label{eq:llfGaussian}
\end{equation}

The gradient or the Hessian of such a log-likelihood are required quite often in statistics in various contexts. Let us supply a few examples. First, Laplace's method (approximation) is very popular technique to approximate stochastic integrals of the form 
\begin{equation}
 \int e^{M l(\pars)}\dint{\pars}
\end{equation}
around a mode $\hat\pars$ of sufficiently smooth negative function $l$ leading to
\begin{equation}
 \int e^{M l(\pars)}\dint{\pars} \to \left(\frac{2\pi}{M}\right)^{\dimpars/2} 
|-H(\hat\pars)|^{-1/2} e^{M l(\hat\pars)}
\end{equation}
as $M(\in \RR^+)\to \infty$ \citep{Wong2001}. The evaluation of Hessian $H(\hat\pars)$ of $l$ is 
needed. Second, when looking for a maximum-likelihood estimator $\mle := \mathop{\rm{argmax}}\limits_{\pars} l(\pars)$ one usually applies some optimization algorithm which requires many evaluations of gradient $\nabla l(\pars)$, such as Conjugate Gradient (CG) or Broyden-Fletcher-Goldfarb-Shanno (BFGS) \citep{Bertsekas1999, Bazaraa2006}.

Third, modern Monte Carlo Markov Chain (MCMC) samplers such as Metropolis-Adjusted Langevin algorithm (MALA) or Hamiltonian Monte Carlo (HMC) also require computing gradients or even Hessians of a log-likelihood with respect to the model parameters for every sample \citep{Brooks2011}.

For any of the above problems, the computation of the derivatives is a key operation and its speedup directly translates to the speedup of the whole algorithm.
For example, in the case of the HMC sampler, the total speedup is roughly proportional to the speedup of the gradient computation of the log-likelihood.

\subsection{Adjoint-state method}
We will employ the \emph{adjoint-state method} (ASM) to efficiently compute the first and the second derivative (gradient and Hessian) of \refe{eq:generallikelihood} with respect to the model parameters $\pars$. 

The ASM is used in many different fields, such as control theory \citep{Lions1971}, data assimilation in meteorology \citep{Lewis2006} or parameter identification \citep{Melicher2013, Cimrak2007}. It is difficult to precisely trace its origin, since it is based on a general principle - the duality. Special dual problems or special test functions in general are used extensively in functional analysis for quite different tasks, e.g. in homogenization theory \citep{Bensoussan1978}. The idea of the 
ASM is to derive a special dual problem to the sensitivity equation of 
\refe{eq:ivp}, which allows one to write the derivative(s) 
of \refe{eq:generallikelihood} 
in a simple form which is inexpensive to evaluate. Usually, one obtains an inner
product(s) in a suitable Hilbert space containing the dual state.

Even if the ASM is a classical method in many different fields, its applications in general statistical literature are rather scarce or could be even considered virtually non-existent. In our opinion, this is due to several reasons.

Mainly, it is a matter of need. Until recently the usual statistical models had only few parameters and the corresponding derivatives were easily evaluable using finite differences. On the other hand, the ASM was mainly used for problems, where derivatives with respect to infinite dimensional parameters are needed, such as the optimal control of partial differential equations (PDEs) \citep{Lions1971}. For those problems, the ASM is the only viable way to compute the gradient of a cost functional. 

The second possible reason is the lack of interdisciplinary publications in statistical literature with the fields where the ASM plays the role of a classical technique. One of the exceptions that elegantly connects the worlds of the PDE-constrained inverse problems and that of Bayesian inference is the paper \citep{Martin2012}. It presents a stochastic Newton method in which MCMC is sampling from a proposal density that builds a local Gaussian approximation based on local derivatives of the log posterior information. The authors exploit adjoint-based gradients and Hessians (as matrix-vector products). They argue that the effective dimension of an parameter estimation problem is often mesh-independent and consequently the Hessian can be approximated by a low-rank approximation computed using a Lanczos process. In general, Meteorology is probably the field where the relationship between differential models and stochastic processes is the most advanced \citep{Lewis2006}.

The third and probably rather influential reason is that the results presented in literature regarding the ASM do not take into account the specifics of statistical estimation, particularly that the measurements can not be altered or interpolated in any way. In this paper, we present an ASM framework for ODE based statistical models, which recognizes and resolves this issue. The ODE case can be addressed in generality, which is not possible for PDE-constrained problems.

The dynamical models described by ODEs are rather widely used in science. They are simply indispensable for acquiring essential knowledge about complex biological systems \citep{Murray2002, Draelants2012} as is the case for other fields studying intricate matters such as psychology and economics. In chemistry, regardless of the criticism \citep{Gillespie1977}, the reaction-rate equations \footnote{Considering spatial phenomena such as diffusion and(or) convection leads to PDE models, see for example \citep{Slodicka2010b}.} are still extensively employed. We are motivated by applications in PK/PD modeling and virology \citep{Lavielle2011, Tornoe2004}.

The sensitivity analysis of ODEs is well established in literature. Let us only mention a classical book on the optimal control of ODEs \citep{Cesari1983}. Moreover,  many results that are intended for PDEs are directly applicable to ODEs, since from the mathematical point of view, an ODE could be simply seen as a PDE without a spatial differential operator. However, as already mentioned, the relevant results presented in literature, do not take into account the specifics of statistical estimation.

The ASM is usually applied in a PDE-constrained context. The fidelity between the data and the PDE-based model is measured in a Lebesgue space $L^p-$norm, particularly in $L^2$ sense, as is also the case of the above mentioned paper \citep{Martin2012}. It implies that the data are considered to be defined almost everywhere in the space or in the space-time in the case of time-dependent problems. This is however in a strong contrast with statistical philosophy. The measurements are ultimately discrete and sacred. E.g. by interpolating the measurements, new ones are generated and that can not be tolerated.

The main contribution of this paper is that it recognizes and resolves this problem. We show, that the discrete data $\meas$ can be combined with the continuous model \refe{eq:ivp} at the level of the likelihood functional \refe{eq:generallikelihood}. The resulting adjoint problem contains a Dirac delta source corresponding to individual measurement times. The developments are fully supported by rigorous proofs.

The subsequent numerical analysis shows that the ASM application for statistical estimation is far from obvious and more work is still needed to reach its efficiency in the deterministic setting.

Moreover, since the ASM is usually applied in an infinite dimensional setting, as explained above, only results for first order derivatives are usually available. For the ODE case, we can supply the Hessian computation as well.

Another contribution is a highly efficient implementation of the results and its benchmarking with respect to finite differences and sensitivity equation approach.

To our best knowledge, we do not know about similar results in the literature.

Last but not least, this interdisciplinary paper aims to popularize this quite underused but potentially very useful method in the statistical community and help those working with ODE based models to compute the corresponding ODE-model sensitivities more efficiently. Recently, with the boom in general availability and dimensionality of data, ODE models with a high number of parameters are being employed. The evaluation of gradients becomes very costly and consequently various derivative-free methods have become more popular, see for example \cite{Delyon1999}. 

One of the domains, where the results presented on the next pages could be particularly appreciated is Systems biology. We refer the interested reader to \cite{Raue2013}, where the
quantitative dynamical modeling is assessed from a rather broad perspective. Notably relevant is a conclusion of the paper that multi-start deterministic parameter optimization using the sensitivity equations (see Section \ref{sec:modelsensitivity} here) for the calculation of derivatives significantly outperforms all other tested algorithms, including a number of stochastic optimization variants which do not make use of derivative information.

As we will show, for certain set-ups, the ASM significantly outperforms the sensitivity equation method for the computation of likelihood derivatives with respect to the parameters of the underlying ODE-model. The gradient can be computed with a cost (measured by model evaluations) that is essentially independent of the number of the parameters. The Hessian can be computed as well, with essentially linear cost in the number of the parameters instead of the quadratic one. Consequently, the use of ASM makes derivative-based (optimization) algorithms for certain setups even more competitive then presented in \cite{Raue2013}.

The paper is structured as follows. In Section 2, we analyze the sensitivity of initial value problems \refe{eq:ivp} with respect to its parameter vector $\pars.$ In Section 3, we present in detail the approach to combine discrete data with a continuous model. Then in Section 4, we obtain the ASM for computing of the gradient and Hessian of \refe{eq:generallikelihood} with respect to $\pars.$ Finally, the implementation is discussed in Section 5 and its efficiency is tested on a number of examples in Section 6.

\section{Sensitivity of model}
\label{sec:modelsensitivity}

In this preparatory section we will discuss the well-posedness of the initial value problem \refe{eq:ivp} as well as the existence of its derivative with respect to the parameters $\pars.$ We follow the presentation in \citep{Zeidler1985} with all the relevant notation, so we can be rather concise. 

The first Gat\`{e}aux differential of a function $f$ with respect to $\bx$ in direction $\bh$ is denoted by $\fd{f}{\bx}{\bh}$. Then, let us denote by $\sv:=\fd{\mv}{\pars}{\bh}$, i.e. the first Gat\`{e}aux differential (we will show it is Fr\'echet as well) of the model function $\mv$ with respect to the parameters $\pars$ in direction $\bh.$ If it exists, the formal differentiation of \refe{eq:ivp} yields that $\sv$ is the solution to the following initial value problem
\begin{equation}
\begin{aligned}
\der{\sv}{t} &= \jac{\mrhs}{\mv}\sv + \jac{\mrhs}{\pars}\bh , 
\quad t\in [0,T], \\ 
\sv(0) &= \jac{\mv_0}{\pars}\bh,
\end{aligned}
\label{eq:sensitivity}
\end{equation}
known as the \emph{sensitivity equation}. Here $\jac{\mrhs}{\pars}: \RR^{\dimpars} \to \RR^\dimmv$ and $\jac{\mrhs}{\mv}: \RR^\dimmv \to \RR^\dimmv$ are the Jacobians of the r.h.s. $\mrhs$ of the model $\refe{eq:ivp}$ with respect to the model parameters $\pars$ and the state variables of the model $\mv$, respectively. Similarly, $\jac{\mv_0}{\pars}$ denotes the Jacobian of the initial value with respect to the model parameters $
\pars$.

Let $\be_i, i=1,\dots, \dimpars$ be the canonical basis in $\RR^\dimpars.$ Solving \refe{eq:sensitivity} for $\bh = \be _i$ for each $i=1,\dots, \dimpars$ yields $\sv = (\pd{\mv_1}{\pars_i}, \pd{\mv_2}{\pars_i},\dots, \pd{\mv_\dimmv}{\pars_i})^t$, if the partial derivatives exist. It means that to compute the whole jacobian $\jac{\mv}{\pars}$ one needs to integrate $\dimpars$ initial value problems \refe{eq:sensitivity}. The complexity is essentially identical to that of the first-order finite difference approximation, as will be confirmed in Section \ref{sec:numerical}. The sensitivity equation approach is however still preferred if high accuracy is needed.

Let us restate the Theorem 4.D from \citep{Zeidler1985} in our context.

\begin{theorem}
Suppose that the mappings $\mrhs:U\subseteq \RR \times \RR^\dimmv \times \RR^\dimpars \to \RR^\dimmv$ and $\mv_0:V\subseteq \RR^\dimpars \to \RR^\dimmv$ are $C^k, k\ge 1$ and that $U$, $V$ are open sets containing $(0, \mv_0(\pars_0), \pars_0)$ and $\pars_0$, respectively. Then:
\begin{itemize}
\item{(a)} There exists an interval $(-a, a), a > 0,$ and an open neighborhood $U(\pars_0)$ such that the initial value problem \refe{eq:ivp} has exactly one solution for each $\pars\in U(\pars_0).$
\item{(b)} The mapping $(t, \pars) \mapsto \mv(t; \pars)$ is $C^k$ on $(-a, a)\times U(\pars_0),$ and \refe{eq:sensitivity} holds.
\end{itemize}
\label{theo:mainTheoremODE}
\end{theorem}

Since our initial value problem \refe{eq:ivp} slightly differs from that of \citep{Zeidler1985} and also for the completeness we present a proof in Appendix \ref{app:proofs}.

From now on, any formal differentiation of $\mv$ with respect to the parameters $\pars$ is justified by Theorem \ref{theo:mainTheoremODE}. The theorem provides only a local result regarding the existence and the uniqueness of the solution $\mv$ of the ivp \refe{eq:ivp}. Consequently, we have to assume that $T<a.$

\section{Connecting the worlds}
\label{sec:connectingtheworlds}
Measurements $\meas_{i}$ are acquired at discrete time points 
$t_{i}$. In statistics, these measurements should not be tampered in any way,
e.g. they cannot be interpolated, which stands for augmentation.

On the other hand the model \refe{eq:ivp} is a continuous one and since 
the adjoint-state method (ASM) deals extensively with the model and the functional 
\refe{eq:generallikelihood}, it is necessary to work in continuous setting. 

We will connect the discrete data and the continuous model on the level of the 
likelihood functional. One 
can write
\begin{equation}
\sum_i \dist(\meas_i, \mf(t_i,\pars)) = \int_0^T \delta\{t-t_i\}\dist(\meas(t),\mf(t, 
\pars)) \dt,
\label{eq:diraccost}
\end{equation}
where, by the classical misuse of notation, $\delta\{t-t_{i}\}$ is the Dirac 
delta function of the set of all measurement times $t_{i}.$ In order to achieve that 
the above integral is well-defined, we will consider a small 
positive $\epsilon$, such that the functions $\meas(t) := \meas_{i}$, 
$t\in(t_{i}-\epsilon, t_i+\epsilon)$ for all measurement times $t_i$ are 
well defined. We emphasize, that by doing so, we do not generate new 
measurements. We merely assume an infinitesimally small interval of their validity. The $\meas(t)$-values outside of intervals $t\in(t_i-\epsilon, 
t_i+\epsilon)$ are irrelevant. For clarity, we extend the function  
$\meas(t)$ outside of these intervals by linear 
interpolation to continuous functions on whole interval $[0,T]$ \footnote{Other continuous ``interpolation'' are possible such as piecewise-linear or by cubic splines, but they are less graphic.}. 

For the well-posedness of \refe{eq:diraccost}, also the model $\mf(t, \pars)$ has to be at least continuous around each $t_i$. Let us assume that $\proj \in \mathcal L(\RR^\dimmv, \RR^\dimo),$ i.e. $\proj$ is a linear operator from $\RR^\dimmv$ to $\RR^\dimo$. The linearity is a sufficient condition for the validity of 
\begin{equation}
\proj(\mv(t, \pars)) - \proj(\mv_0(\pars)) = \int_0^t \proj(\mrhs(t, \pars, \mv)) \dt,
\end{equation}
which will be needed for the subsequent developments. From now on we write $\proj \mv$ instead of $\proj(\mv)$. Let us point out that this assumption is usually not restrictive in practical applications. The eventual non-linear transformations can be applied a priori to the data $\meas$ or included in $\mrhs.$ Now, since the solution $\mv$ to \refe{eq:ivp} is at least continuously differentiable (Theorem \ref{theo:mainTheoremODE}) and a linear operator preserves continuity, $\mf(t, \pars)$ is trivially continuous.

As a convenience, for any distribution $d$ and sufficiently smooth function 
$f$, we denote by $\dual{d}{f}$ the duality between them on the time interval 
$[0,T]$. We will also need the scalar product $\scal{\cdot}{\cdot}$ in the 
Hilbert space $L^2([0,T]).$ Using this notation, \refe{eq:diraccost} can be 
rewritten as
\begin{equation}
\int_0^T \delta\{t-t_i\}\dist(\meas(t),\mf(t, 
\pars)) \dt{t} = \dual{\delta\{t-t_i\}}{\dist(\meas(t),\mf(t, 
\pars))}.
\label{eq:diraccost2}
\end{equation}
Let us introduce short notation $\{\delta\}$ for $\delta\{t-t_i\}$.
At last, the equality \refe{eq:diraccost} defines a seminorm on $C([0,T], \RR ^\dimo)$, since the l.h.s. is a discrete norm. We denote this seminorm simply as 
$\norma{\cdot}$.

Using the gluing notation above, we can prove the following lemma that allows us to evaluate the first differential of \refe{eq:generallikelihood} using the solution $\sv$ to the sensitivity equation \refe{eq:sensitivity}.

\begin{lemma}
Let the assumptions of Theorem \ref{theo:mainTheoremODE} be fulfilled for $k=1$ and let the metric $\dist$ be $C^1.$ Then the functional \refe{eq:generallikelihood} is Fr\'echet differentiable and the differential $\fd{\llf}{\pars}{\var\pars}$ can be expressed as
\begin{equation}
\fd{\llf}{\pars}{\var\pars} =  
\dualk{\{\delta\}\dist_{\mv}(\meas,\mf(t, 
\pars))}{\sv},
\label{eq:rfirstder}
\end{equation}
where $\sv$ is the unique solution to sensitivity equation \refe{eq:sensitivity}.
\label{lemma:gradSensitivity}
\end{lemma}
A proof can be found in Appendix \ref{app:proofs}. Moreover, due to the linearity of \refe{eq:sensitivity}, the differential $\fd{\llf}{\pars}{\var\pars}$ can be easily written in the linearized form $\fd{\llf}{\pars}{\var\pars} = \llf'(\pars) \var\pars.$ Since we work in finite dimensional spaces, the expression
\begin{equation}
\nabla \llf(\pars)\cdot \var\pars := \llf'(\pars) \var\pars \quad \mbox{for all }\var\pars\in\RR ^\dimpars
\label{eq:gradient}
\end{equation}
well defines the gradient $\nabla \llf(\pars)$ of $\llf$ as an element in $\RR ^\dimpars$ for each fixed $\pars,$ i.e. $\nabla \llf: \RR ^\dimpars \to \RR ^\dimpars.$

As explained in Section \ref{sec:modelsensitivity}, $\dimpars$ initial value problems \refe{eq:sensitivity} have to be computed to evaluate $\nabla \llf(\pars)$ for some $\pars.$

\section{Adjoint-state method}
\label{sec:adjoint}

In this Section we will introduce the \emph{adjoint-state method} (ASM) for computation of the gradient and the Hessian of \refe{eq:generallikelihood}. The results are strongly influenced by the peculiar coupling between the discrete measurements and the continuous model \refe{eq:ivp}. Let us directly present the main statement.
\begin{theorem} Let the assumptions of Lemma \ref{lemma:gradSensitivity} be fulfilled. Then the first Fr\'echet differential in \refe{eq:rfirstder} can be also written as 
\begin{equation}
\fd{\llf}{\pars}{\var\pars} =  -\dv^t(0)\jac{\mv_0}{\pars}\bh-\scal{\jac{\mrhs}{\pars}\var{
\pars}}{\dv}
\label{eq:rfirstderAdjoint}
\end{equation}
where $\dv$ is the unique solution to the following initial value problem
\begin{equation}
\begin{aligned}
\der{\dv}{t} &= -\jact{\mrhs}{\mv}\dv + \{\delta\} \dist_{\mv}(\meas,\mf(t, 
\pars)),
\quad t\in [0,T], \\ 
\dv(T) &= 0.
\end{aligned}
\label{eq:adjoint}
\end{equation}
\label{theo:gradAdjoint}
\end{theorem}
A proof is again presented in Appendix \ref{app:proofs}. Obviously, Equation~\refe{eq:rfirstderAdjoint} is written in linearized form. We get
\begin{equation}
\nabla \llf = -\dv^t(0)\jac{\mv_0}{\pars}-\scal{\dv}{\jac{\mrhs}{\pars}},
\label{eq:gradAdjoint}
\end{equation}
where the second term on the r.h.s. is a vectorial integral. This is a very efficient way to compute the gradient. One has to only integrate one adjoint problem \refe{eq:adjoint} and evaluate the expression \refe{eq:gradAdjoint}, i.e. to compute $\dimpars$ scalar products in $L^2([0,T])$.

Let us discuss the result a little. The ivp \refe{eq:adjoint} is a special ODE. First, it has absolutely no physical, chemical, biological or any other interpretation of the underlying scientific field of equation~\refe{eq:ivp}. The best way to look at it is that it is a special dual problem (see the proof) to the sensitivity equation \refe{eq:sensitivity}, which allows us to efficiently compute the gradient of \refe{eq:generallikelihood} (and the Hessian as well as we will see.) Then, it is a final time problem to be integrated from $T$ to the initial time $0.$ It is a linear ODE like the sensitivity equation. Its r.h.s. contains the term $\{\delta\} \dist_{\mv}(\meas,\mf(t, \pars))$, which expresses how quickly the distance between the data and the model changes when changing the model variable $\mv.$

Probably the most important fact to note about the ASM is that it operates at a higher level than the sensitivity equation method. It does not supply the derivative of the state $\mv$, but directly the one of the likelihood functional \refe{eq:generallikelihood}. By considering the model together with \refe{eq:generallikelihood}, the efficiency can be achieved.

The numerical issues regarding the integration of \refe{eq:adjoint} will be discussed in Section \ref{sec:ASMimplementation}.

\begin{example}
Let us consider the distance \refe{eq:l2weightedMetric} corresponding to the multivariate normal distribution of the residual errors. Then the derivative $\dist_{\mv}(\meas,\mf(t, \pars))$ in the r.h.s of \refe{eq:adjoint} reads
\begin{equation}
\dist_{\mv}(\meas,\mf(t, \pars)) = -\proj^t\recov^{-1}(\meas_i-\mf(t_i, \pars)).
\end{equation}
The adjoint problem is dependent on the residual covariance matrix $\recov$ and on the post-processing operator $\proj.$

\end{example}

\subsection{Evaluating Hessian}
\label{sec:Hessian}

Let us depict the second Gat\`{e}aux differential of a functional $f$ with respect to $\bx$ in directions $\bh_1$ and $\bh _2$ as $\sd{f}{\bx}{\bh_1, \bh_2}$. Then, let us introduce notation $\ssv := 
\sd{\mv}{\pars}{\bh_1, \bh_2}.$ We will show that $\ssv$ is Fr\'echet as well. By formally differentiating \refe{eq:sensitivity} one more time with respect to $\pars$ we obtain that $\ssv$ is a solution to the following initial value problem 
\begin{equation}
\begin{aligned}
\der{\ssv}{t} &= \pddl{\mrhs}{\pars\pars}\bh_1\bh_2 
+ \pddl{\mrhs}{\pars\mv}\bh_1\sv_2 + 
\pddl{\mrhs}{\mv\pars}\sv_1\bh_2\\
& + \pddl{\mrhs}{\mv\mv}\sv_1\sv_2 + \jac{\mrhs}{\mv}\ssv,\quad t\in 
[0,T], \\ 
\ssv(0) &= \pddl{(\mv_0)}{\pars\pars}\bh_1\bh_2
\end{aligned}
\label{eq:secondsens}
\end{equation}
known as the \emph{second sensitivity equation}. Here $\sv_1,$ $\sv_2$ are the solutions to \refe{eq:sensitivity} for $\bh=\bh _1,$ $\bh=\bh _2,$ respectively. The second order derivatives in \refe{eq:secondsens} are essentially three-dimensional tensors. In Appendix \ref{app:proofs} the following lemma is proven.

\begin{lemma}
Let the assumptions of Theorem \ref{theo:mainTheoremODE} be fulfilled for $k=2$ and let the metric $\dist$ be $C^2.$ Then the second Fr\'echet differential of  \refe{eq:generallikelihood} with respect to the model parameters $\pars$ can be written as
\begin{equation}
\sd{\llf}{\pars}{\var\pars_1, \var\pars_2} = 
\dual{\{\delta\}}{\dist^2_{\mv\mv}\sv_1\sv_2}
+\dual{\ssv}{\{\delta\}\dist_{\mv}(\meas,\mf(t, \pars))},
\label{eq:rsecder}
\end{equation}
where $\ssv$ is the unique solution to \refe{eq:secondsens}. Moreover, the second term in \refe{eq:rsecder} can be rewritten using the solution $\dv$ to \refe{eq:adjoint} as
\begin{equation}
\begin{aligned}
\dual{\ssv}{\{\delta\}\dist_{\mv}(\meas,\mf(t, \pars))}
&= -\pddl{(\mv_0)}{\pars\pars}\bh_1\bh_2\cdot\dv(0) -\scal{\pddl{\mrhs}{\pars\pars}\bh_1\bh_2}{\dv}\\
&-\scal{\pddl{\mrhs}{\pars\mv}\bh_1\sv_2}{\dv}
-\scal{\pddl{\mrhs}{\mv\pars}\sv_1\bh_2}{\dv}
-\scal{\pddl{\mrhs}{\mv\mv}\sv_1\sv_2}{\dv}.
\label{eq:SecDerSens}
\end{aligned}
\end{equation}
\label{lemma:HessianAdjoint}
\end{lemma}

Solving \refe{eq:secondsens} for $\bh_1 = \be _i$, $\bh _2 = \be _j$ for each $i=1,\dots, \dimpars,$ $i=1,\dots, \dimpars$ yields $\ssv = (\pddm{\mv_1}{\pars_i\partial\pars_j},$ $ \pddm{\mv_2}{\pars_i\partial\pars_j},\dots, \pddm{\mv_\dimmv}{\pars_i\partial\pars_j})^t$. It means that to compute the Hessian $H_{\pars}(\mv)$, one needs to integrate the second sensitivity equation \refe{eq:secondsens} $p(p+1)/2$ times. For that one moreover needs to compute $\dimpars$ sensitivities $\sv_i$ for each $h=\be _i,$ $i=1,\dots, \dimpars$. The cost is essentially identical to that of the first order finite difference approximation. Again, it is beneficial if high accuracy is needed.

On the other hand, the evaluation of the Hessian $H_{\pars}\llf$ of \refe{eq:generallikelihood} via \refe{eq:SecDerSens} requires us to only
compute one adjoint problem \refe{eq:adjoint}, $\dimpars$ sensitivity 
equations \refe{eq:sensitivity} and $p(p+1)/2$ times the four scalar products from 
\refe{eq:SecDerSens}. This is a very efficient and accurate way how to compute the Hessian.

As before with the gradient, we see that the ASM supplies the sensitivity at the level of the functional, not that of the underlying model state $\mv.$

\subsubsection{Hessian via adjoint with finite differences}

Let us present an alternative way to efficiently compute the Hessian of \refe{eq:generallikelihood}, which is slightly less accurate than using \refe{eq:SecDerSens} but much simpler to implement. The idea is to combine \refe{eq:rfirstderAdjoint} with finite differences as follows
\begin{equation}
  H_i(\llf(\pars)) \approx \frac{\nabla \llf(\pars + h \be_i) - \nabla \llf(\pars)}{h},
\label{eq:FDadjoint}
\end{equation}
where $H_i$ stands for the i-$th$ column of $H$ (or row) and $h$ is a small positive real number. We recall that $\{\be_i: 1 \le i \le \dimpars\}$ is the canonical basis in $\RR^{\dimpars}.$ Each of $\dimpars$ gradients $\nabla \llf(\pars + h \be_i),$ $1 \le i \le \dimpars$ is computed using \refe{eq:rfirstderAdjoint}. Together $p+1$
adjoint initial value problems \refe{eq:adjoint} need to be integrated.

\section{Implementation of ASM}
\label{sec:ASMimplementation}

In this Section we will describe an implementation of the ASM presented in Section \ref{sec:adjoint}. At the core of the developments is the adjoint initial value problem \refe{eq:adjoint}. Even if it is a rather simple linear ODE-system, it is a quite difficult one to solve numerically because of its r.h.s. containing the Dirac delta function source term. 

Our implementation closely follows the constructive proof of Theorem \ref{theo:gradAdjoint} in Appendix \ref{app:proofs}. At each measurement time $t_i$, ODE-solver is stopped, $\dist_{\mv}(\meas_i,\mf(t_i, \pars))$ is explicitly added to the current solution and then the integration is resumed. We solve a sequence of initial value problems \refe{eq:adjointWithoutDirac} instead of the original ivp \refe{eq:adjoint}. 

Unfortunately, the repetitive restarting of the ODE solver has a negative impact on the performance. The derivative $\dist_{\mv}(\meas_i,\mf(t_i, \pars))$ is added to the dynamical system at once via the initial condition $\dv(t_{i})$ and since the time derivative $\der{\dv}{t}$ from \refe{eq:adjointWithoutDirac} is proportional to $\dv(t_i)$, it encounters a jump at each measurement point. Consequently, the steepness of the solution forces the ODE-solver to advance in many small time steps, which leads to a high number of iterations. We will see in Section 6, that the efficiency of ASM is indeed strongly dependent on the number of measurements.

However, equation~\refe{eq:adjointWithoutDirac} is a quite simple linear ODE-system, which should be exploitable in multiple ways. Although increasing the numerical efficiency of backward integration while preserving the statistical rigor is a very interesting scientific goal, it is out of the scope of this contribution and left for the future research.

We tackle \refe{eq:adjoint} using CVODES solver from the SUNDIALS package \citep{Hindmarsh2005}. CVODES is an extension of the CVODE code with both forward and backward sensitivity abilities \citep{Serban2005}.

The numerical experiments presented in Section \ref{sec:numerical} are computed in 
\hfill \diffmem\ \citep{Haber2016}. It is a package for the fitting of mixed-effect models
constrained by differential equations. The package is under 
active development by the authors and the algorithms presented in this paper are only 
a subset of its abilities.

At present, only ODE dynamical models are supported. \diffmem{ } employs the ODE solvers of CVODE for quick and robust solution of those models. It uses Eigen linear 
algebra package (library) to represent its internal memory containers and to 
solve underlying linear systems.

\begin{remark}
Let us imagine, we would not explicitly integrate the Dirac delta function out. It can be approximated in many different ways, but the most suitable from the statistical point of view (owing to the central limit theorem) is the approximation using Gaussian
\begin{equation}
\delta_i^{\sigma}(t) := \frac{1}{\sigma\sqrt{2\pi}}
e^{-\frac{(t-t_i)^2}{2\sigma^2}},
\label{eq:diracsigma}
\end{equation}
where $\sigma$ can be seen as a measure of the trust that the measurements have been taken precisely at the times $t_i.$ Let us define $\delta^{\sigma}(t) := \sum_i \delta_i^{\sigma}(t).$ Using this approximation, the adjoint system \refe{eq:adjoint}
becomes 
\begin{equation}
\begin{aligned}
\der{\dv}{t} &= -\jact{\mrhs}{\mv}\dv + \delta^{\sigma}(t)\dist_{\mv}(\meas,\mf(t, 
\pars)), 
\quad t\in [0,T], \\ 
\dv(T) &= 0.
\end{aligned}
\label{eq:adjointsigma}
\end{equation}

Here, the adjoint problem \refe{eq:adjointsigma} represents an interesting antagonism between the efficiency of the ASM and the statistical rigour one expects.
The higher the trust in the measurement times, the smaller the $\sigma$ and consequently higher the derivative of the r.h.s of \refe{eq:adjointsigma} with respect to time which makes this dynamical system more and more difficult for an ODE solver to integrate.\footnote{The variance $\sigma^2$ has here a purely ad hoc use for the above argument. We are not interested if it is prescribed or estimated from the data.}
\end{remark}

\section{Numerical experiments}
\label{sec:numerical}
In this section we will consider for simplicity but without any loss of generality the Gaussian log-likelihood \refe{eq:llfGaussian} with $\recov = I.$

For all the experiments, the ODE solvers' absolute accuracies are set to $10^{-14}$ and the relative ones to $10^{-10}.$ These accuracies are sufficient to remove considerations about accuracy of the ODE-solver from the analysis.

We will study the efficiency and robustness of the adjoint-state method (ASM) for computing the derivatives of the likelihood.

\subsection{Linear model}
\label{sec:linearModel}

To start, let us consider the classical linear ordinary differential equation 
(ODE)
\begin{equation}
\begin{aligned}
\der{\mv}{t} &= A\mv, \quad t\in [0,T], \\ 
\mv(0) &= \mv_0,
\end{aligned}
\label{eq:linear_model}
\end{equation}
where $A$ is a $d\times d-$dimensional matrix, the elements of which represent the model parameters. This simple model is ideal toy-example to comprehend the importance of ASM for models with high dimensional parametric space.

Let us consider diagonal matrix $A$. Then the dimensions of the parametric space and of the solution coincide ($\dimpars = \dimmv$). Moreover, we can easily calculate the exact solution
\begin{equation}
\mv_i = \mv_{0,i} e^{\pars_i t},\quad i=1, \dots, \dimpars,
\end{equation}
where $\pars_i = A_{ii}.$ 

We consider 13 different dimensions of $\pars,$ ranging from $2$ to $122$. For each of them we have randomly generated $100$ parameter-samples $\pars$  as follows 
\begin{equation}
 \pars_i \sim U[-1.1, -0.1],\quad 1 \le i \le \dimpars.
\label{eq:ipartgeneration}
\end{equation}
We set $\mv_0 = \mathbf{1}.$ The number of 
observation time points $\nom$, regularly spread in $[0,100]$, is set constant to $11$.

The corresponding synthetic data $\meas$ are also perturbed as follows:
\begin{equation}
 \meas_i \sim U[\meas_i, \meas_i + 10^{-1}\max(\meas)],\quad 1 \le i \le \nom,
\label{eq:datageneration}
\end{equation}
where $\max(\meas)\in\RR^\dimo$ is a constant vector containing at each position the same maximum. We would like to emphasize that any reasonable perturbation leads to the same results. It is only important that the data are perturbed outside of the log-likelihood mode.

We have computed the gradient of likelihood using finite 
differences (FD), the ASM approach \refe{eq:rfirstderAdjoint} and using the 
sensitivity equation (SE) \refe{eq:sensitivity}. We recall that the last two 
approaches are implemented using the CVODES forward- and backward- sensitivity 
abilities and all the common settings are identical to make comparison 
as sound as possible. The results are presented in Figure \ref{fig:gradientDLM}.

\def\multfig{1.}
\begin{figure}
\includegraphics[width=\multfig\textwidth]{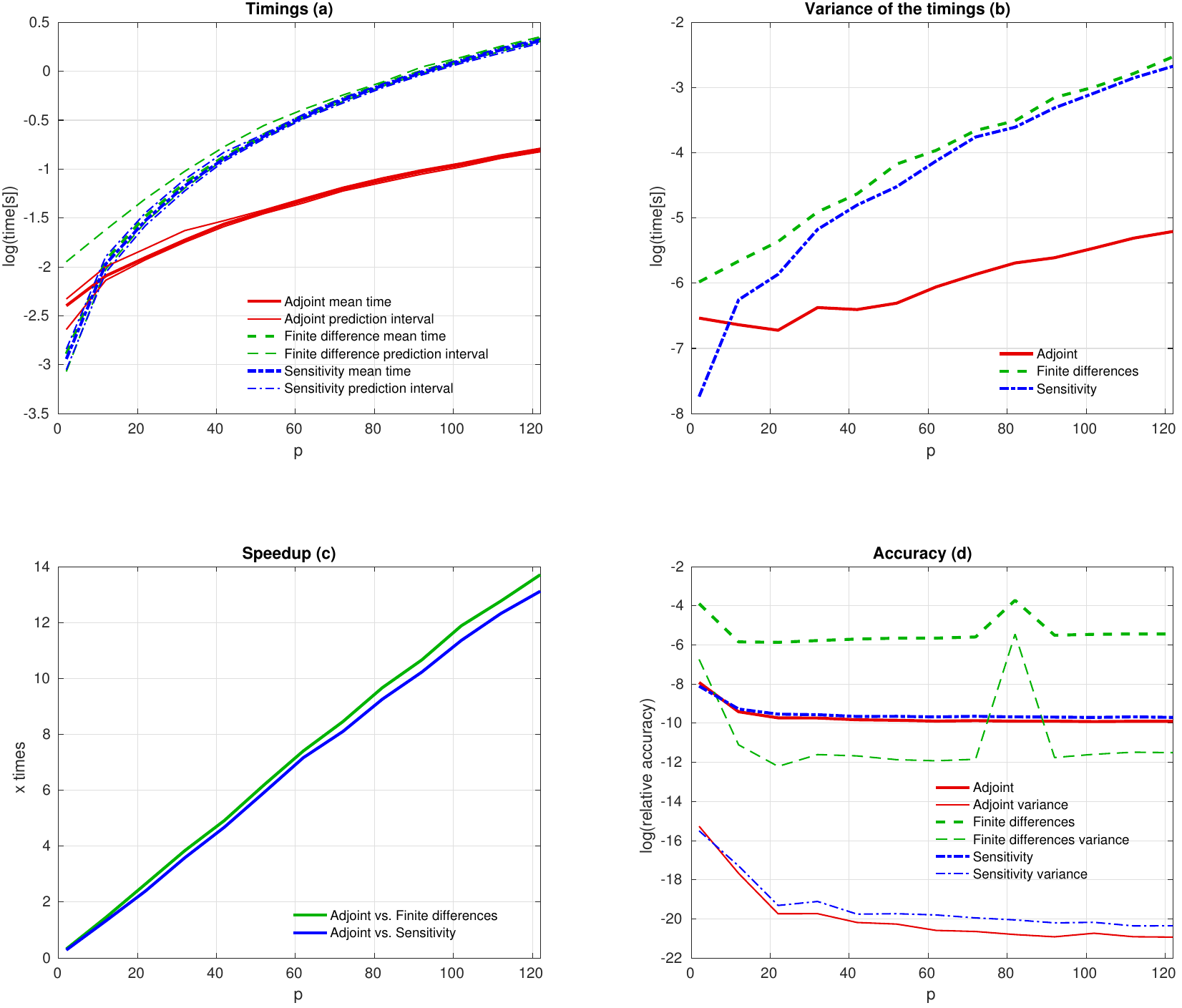}
\caption{Linear ODE model: comparison of different methods for log-likelihood gradient computation with respect to the growing parameter space dimension of the ODE model}
\label{fig:gradientDLM}
\end{figure}

The timings of FD, ASM and of SE are presented in Figure 
\ref{fig:gradientDLM}(a). The first conclusion is that our implementation of 
SE-approach is optimal since the timings of FD and SE more or less coincide. 
Actually SE is always a slightly faster method. Given the significantly higher 
accuracy of SE with respect to FD (\ref{fig:gradientDLM}(d)), it renders 
FD-approach redundant.

Somewhere around $10$ parameters ASM becomes on average more efficient than SE. 
Moreover, given the non-parametric prediction intervals based on the $100$ 
samples, it is from around $15$ parameters virtually always more 
time-efficient than SE. This reasoning is conservative since it does not take the 
correlation between ASM and SE timings into account. Moreover, the variance of 
timings is for $\dim(\pars) > 10$ lower for ASM than for SE, see Figure 
\ref{fig:gradientDLM}(b), rendering timing predictions for ASM more reliable.

The time efficiency of both ASM and SE is negatively impacted by exclusive use 
of dense matrices in \diffmem. The equations \refe{eq:sensitivity}, 
\refe{eq:adjoint}, \refe{eq:rfirstderAdjoint} require evaluation of Jacobians 
$\jac{\mrhs}{\mv}$ and $\jac{\mrhs}{\pars}$. These are extremely sparse 
\footnote{The diagonal system \refe{eq:linear_model} is the most sparse system 
one can think of.}. More importantly, because of the dense matrix 
implementation only rather small systems can be solved. Sparse matrix
implementation is planned for the future versions of \diffmem. Both ASM and SE 
are influenced to the same degree and the relative comparison holds.

The speedup of ASM vs. SE (\ref{fig:gradientDLM}(c)) is roughly linear in 
the number of the parameters but it slows down slightly for higher parameter 
dimensions. The suspected cause here is the cost of memory access when CVODES 
evaluates the forward solution $\mv$ during the backwards integration of 
\refe{eq:adjoint}.

The accuracy of both ASM and SE with respect to the exact gradient of the 
likelihood \refe{eq:llfGaussian} is presented in Figure \ref{fig:gradientDLM}(d). Both 
methods achieve virtually identical accuracy since the forward- and backward- 
solvers use the same relative and absolute tolerances.

\begin{figure}
\includegraphics[width=\multfig\textwidth]{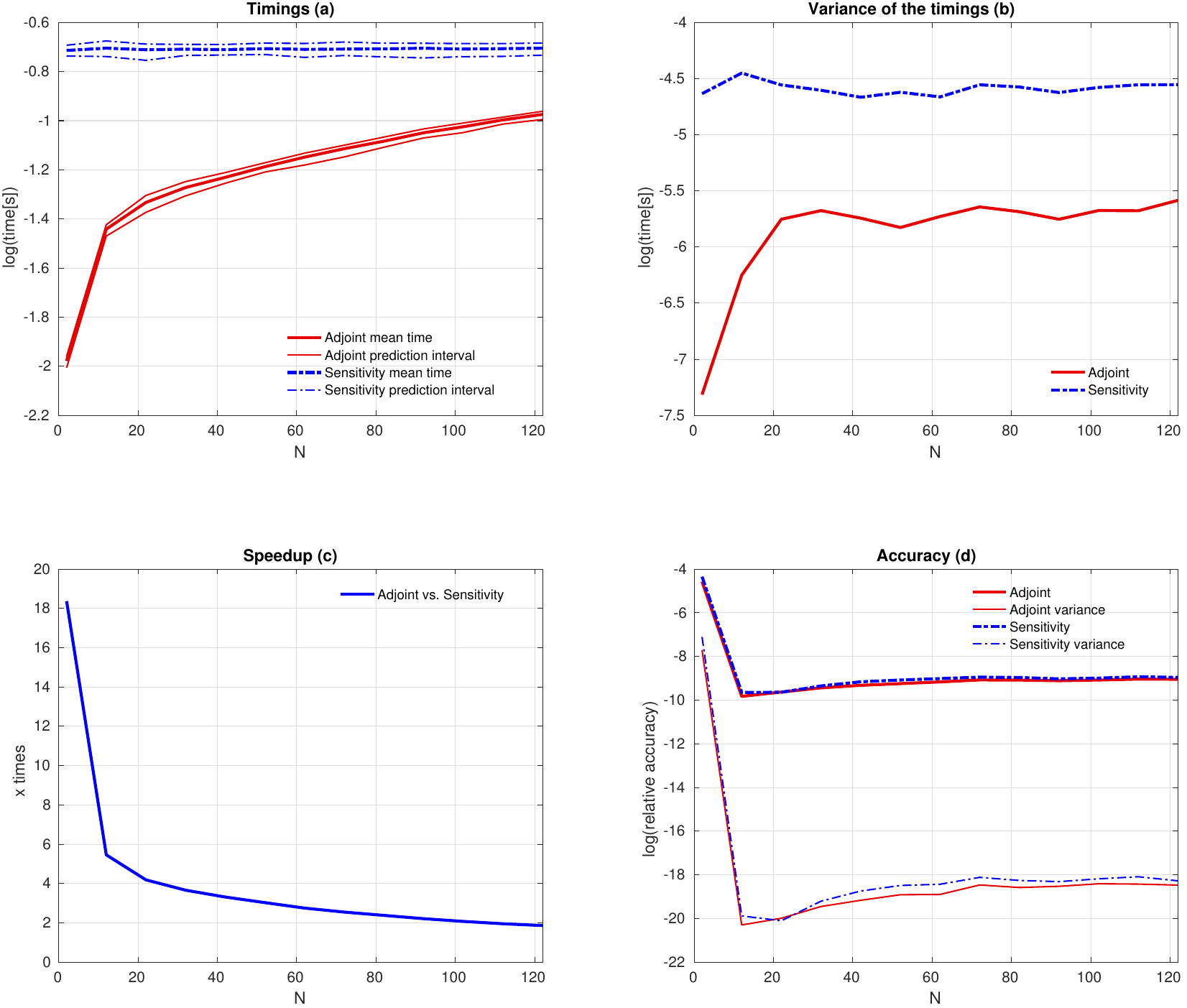}
\caption{Linear ODE model: comparison of different methods for log-likelihood gradient computation with respect to the growing number of the number of time observations}
\label{fig:gradientDLMnoto}
\end{figure}

Now we will examine the efficiency of ASM and SE with respect to the number of 
time observations. We fix the dimension of the problem at e.g. $\dim(\pars) = 
50.$ The number of time observations $\dim(y)$ in $[0, 100]$ fluctuates 
between $2$ and $122$ in $12$ steps. For each number we again compute $100$ 
gradients using \refe{eq:sensitivity} and \refe{eq:rfirstderAdjoint}. The 
parameters $\pars$ are again generated using \refe{eq:ipartgeneration}. The 
results are presented in Figure \ref{fig:gradientDLMnoto}.

The SE approach efficiency is essentially independent of the number of time 
observations (\ref{fig:gradientDLMnoto}(a).) The ASM efficiency however 
decreases with increasing number of observations. As explained in Section 
\ref{sec:ASMimplementation}, the backward adjoint integrator needs after each 
data point a large number of small time steps to account for the steepness of 
the adjoint solution $\dv.$  The negative impact is the most clearly visible in 
Figure \ref{fig:gradientDLMnoto}(c). For many practically relevant problems \footnote{PK/PD, virology.}, the number of measurements is rather low, making this issue less pronounced. Anyhow, increasing the numerical efficiency of backward integration while preserving the statistical rigor is obviously a very interesting direction for future 
research.

Implicitly, since for the diagonal linear model $p = n$, also the dimension of the solution space plays a role. But we do not compare the speed and accuracy of the different methods with respect to $\dimmv$ or $\dimo$, since their complexities with respect to these are the same.

Now, again using the problem \refe{eq:linear_model}, we will illustrate the 
efficiency of computing the Hessian of \refe{eq:llfGaussian} with respect to $\pars$ 
employing the expression \refe{eq:SecDerSens}. We compare this (SA approach) 
with Hessian evaluated using the finite difference approximation (FD) and the 
one computed using \refe{eq:FDadjoint} (FA).

\begin{remark}
First-order Gauss-Newton approximation of the Hessian, where the second term in \refe{eq:rsecder} is neglected, is not included in the comparison. When the model does not yet well approximate the data, the second order term \refe{eq:SecDerSens} can be arbitrarily large with 
respect to the first order term in \refe{eq:rsecder}. This is a well known fact but often overlooked. Let us return to the linear model \refe{eq:linear_model} for a deeper insight. In this case, the first order approximation $F$ of the Hessian of the likelihood $H$ is
\begin{equation}
  \begin{aligned}
    F_{k,k} &= -\sum_{i=1}^{\nom} e^{\pars_k t_i}e^{\pars_k t_i} 
t_i^2, \quad k=1,\dots,\dimpars\\
    F_{k,l} &= 0, \quad k\neq l
  \end{aligned}
\end{equation} and the second order term $S$ is 
\begin{equation}
  \begin{aligned}
  S_{k,k} &= -\sum_i^{\nom} \left(e^{\pars_k t_i} - 
\meas_i\right)e^{\pars_k t_i} t_i^2,   \quad k=1,\dots,\dimpars\\
  S_{k,l} &= 0 , \quad k\neq l.
  \end{aligned}
\end{equation}
We see that the first order approximation $F$ carries absolutely no information about how far the solution is from the data. The Gauss-Newton approximation error can thus be arbitrarily large when $\meas$ is not well approximated by the solution $e^{\pars t},$ especially for the values corresponding to small measurement times.

This is for example exploited in the well-known Levenberg-Marquardt method for the least-square minimization \citep{More1978}, which dynamically switches from the gradient descent method (GD) to the Gauss-Newton (GN) method. The GD is used to get sufficiently close to a minimum, so that the GN approximation is reliable.
\end{remark}

\begin{figure}
\includegraphics[width=\multfig\textwidth]{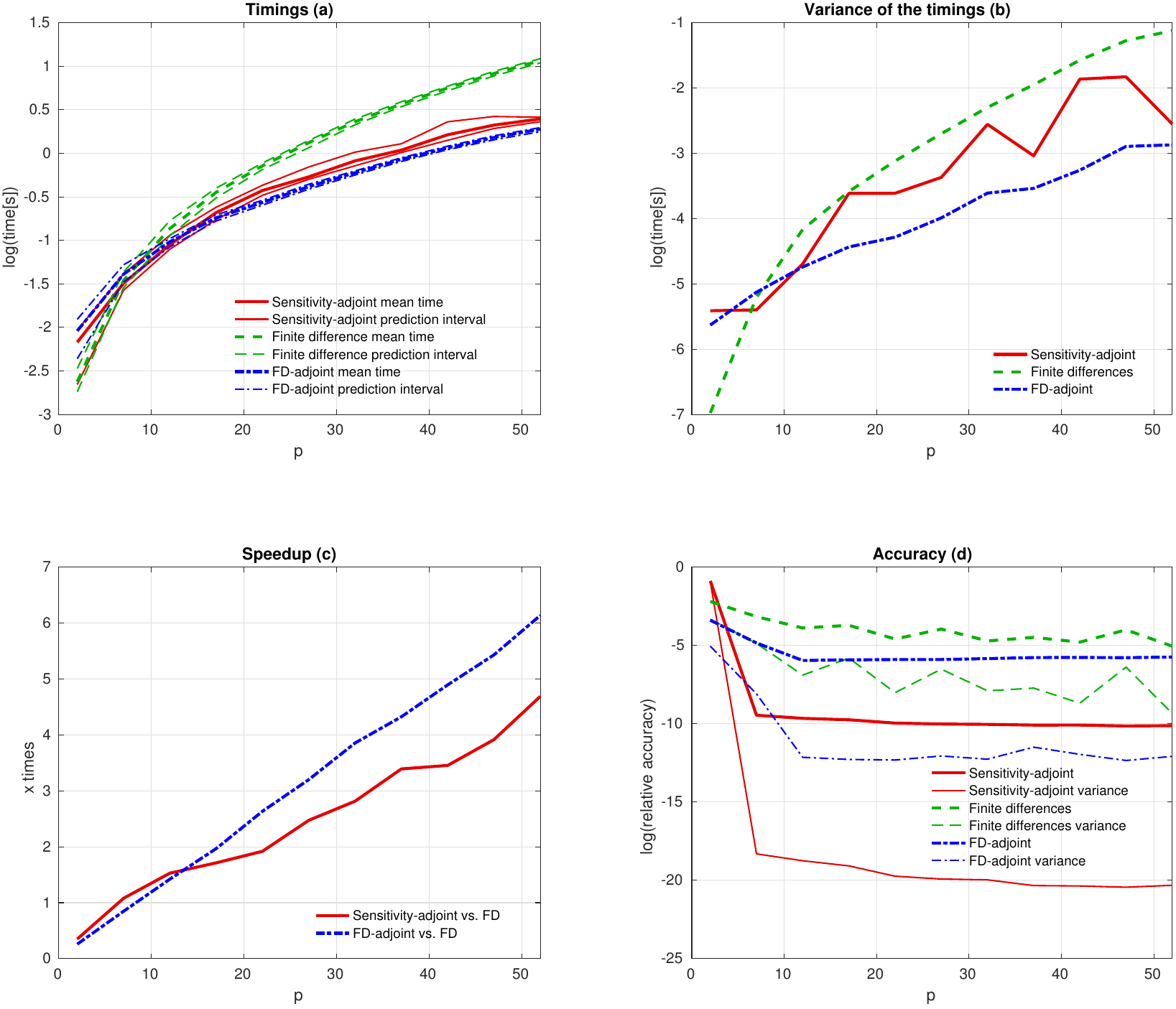}
\caption{Linear ODE model: comparison of different methods for log-likelihood Hessian computation with respect to the growing parameter space dimension of the ODE model}
\label{fig:hessianDLM}
\end{figure}

The overall setup stays identical to the one used for the gradient, i.e. the one corresponding to Figure \ref{fig:gradientDLM}. For convenience, we consider a shorter parameter range $\dim(\pars)\in[2,52], $ since the finite difference approximation of Hessian, to which we compare the ASM, has quadratic complexity in $\dimpars$. It makes the experiments more time prohibiting in comparison to the gradient. The results are depicted in Figure \ref{fig:hessianDLM}.

FD-adjoint, i.e. approximating Hessians using \refe{eq:FDadjoint}, is the most 
time-efficient approach (Figure \ref{fig:hessianDLM}(a)). The speedup with respect 
to the finite difference approximation (FD) is linear in $\dim(\pars)$ as expected 
(Figure \ref{fig:hessianDLM}(c)). The FD-adjoint Hessian accuracy is usually 
sufficient (Figure \ref{fig:hessianDLM}(d)) and moreover it is behaving well
as a function of the dimension of the parametric space.

If higher accuracy up to machine precision is desirable, one can compute 
Hessian using \refe{eq:SecDerSens}, i.e. using SA-approach. Our 
SA-implementation is however clearly slower than the FD-adjoint despite of 
a rather optimal coding. FD-adjoint is superior time-wise mainly due to its 
simplicity.

To conserve space, we do not include any experiments for Hessian with respect to the number of measurement times $\nom$. But obviously, for the Hessian computed via FD-adjoint, the results for gradient are directly applicable. We will analyze the dependence on $\nom$ for the following model in Section \ref{section:HIV}. Here we have focused on $\dimpars-$ scaling only, which cannot be tested for the realistic model.
 
\subsection{Latent dynamic HIV model}
\label{section:HIV}

Now we are going to assess the efficiency and accuracy of ASM for a more 
realistic model - latent dynamic HIV model from \cite{Lavielle2011}:
\begin{equation}
  \begin{aligned}
    \der{\tni}{t} &=\lambda - (1-\enrti)\gamma\tni\vi - \muni\tni,\\
    \der{\tl}{t}  &= (1-\pi)(1-\enrti)\gamma \tni\vi - \al\tl - \mul \tl,\\
    \der{\ta}{t}  &= \pi(1 - \enrti)\gamma \tni \vi + \al\tl -\mua\ta,\\
    \der{\vi}{t}  &= (1-\epi) p\ta - \muv \vi,\\
    \der{\vni}{t} &= \epi p\ta - \muv \vni,
  \end{aligned}
 \label{eq:HIVmodel}
 \end{equation}
where $\tni$ is the number of not-infected CD4 cells, $\tl$ of latent infected 
CD4 cells and $\ta$ the number of active infected CD4 cells producing new 
virons.  The number of infectious viruses is $\vi$ and the non-infectious 
$\vni$. The 11 parameters $\pars$  represent mostly rates of 
change. The two of them $\enrti, \epi \in [0,1]$ represent the efficacies of two types of 
antiviral therapies. The available measurements $\meas_i$ are restricted to the 
cumulative counts of CD4 cells and the virons, i.e. $V_{ij} = \vi + \vni$ and 
$T_{ij} = \tni + \tl + \ta$ respectively. For the details see 
\cite{Lavielle2011}. We have $\dimpars = 11,$ $\dimmv = 5$ and $\dimo = 2.$

The setup of the experiments stays rather similar to the previous ones. The parameters are generated randomly around $\pars_0$ which is presented in Table \ref{table:HIVparameters} as follows:
\begin{equation}
 \pars_i \sim U[0.95\pars_{0,i}, 1.05\pars_{0,i}],\quad 1 \le i \le \dimpars.
\label{eq:ipartgeneration2}
\end{equation}
The efficacies $\enrti$ and $\epi$ can be sometimes generated out of the allowed range $[0, 1).$ We project them back:
\begin{equation*}
\pars_i = \min(\pars_i, 0.999), \quad i \in \{10, 11\}.
\end{equation*}

The corresponding synthetic measurements $\meas$ are perturbed using \refe{eq:datageneration}. 

\begin{table}
\caption{\label{table:HIVparameters}Parameters of the latent dynamic HIV model}
\resizebox{\columnwidth}{!}{%
\fbox{
\begin{tabular}{c|*{11}{c}}
$\pars$ & $\lambda$ & $\gamma$ & $\muni$ & $\mul$ & $\mua$ & $\muv$ & $p$ & 
$\al$ & $\pi$  & $\enrti$ & $\epi$\\
\hline
$\pars_0$ & $2.61$ & $.0021$ & $.0085$ & $.0092$ & $.289$ & $30$ & $641$ &
$\num{1.6e-5}$ & $.443$ & $.90$ & $.99$\\ 
\end{tabular}}}
\end{table}

\begin{figure}
\includegraphics[width=\multfig\textwidth]{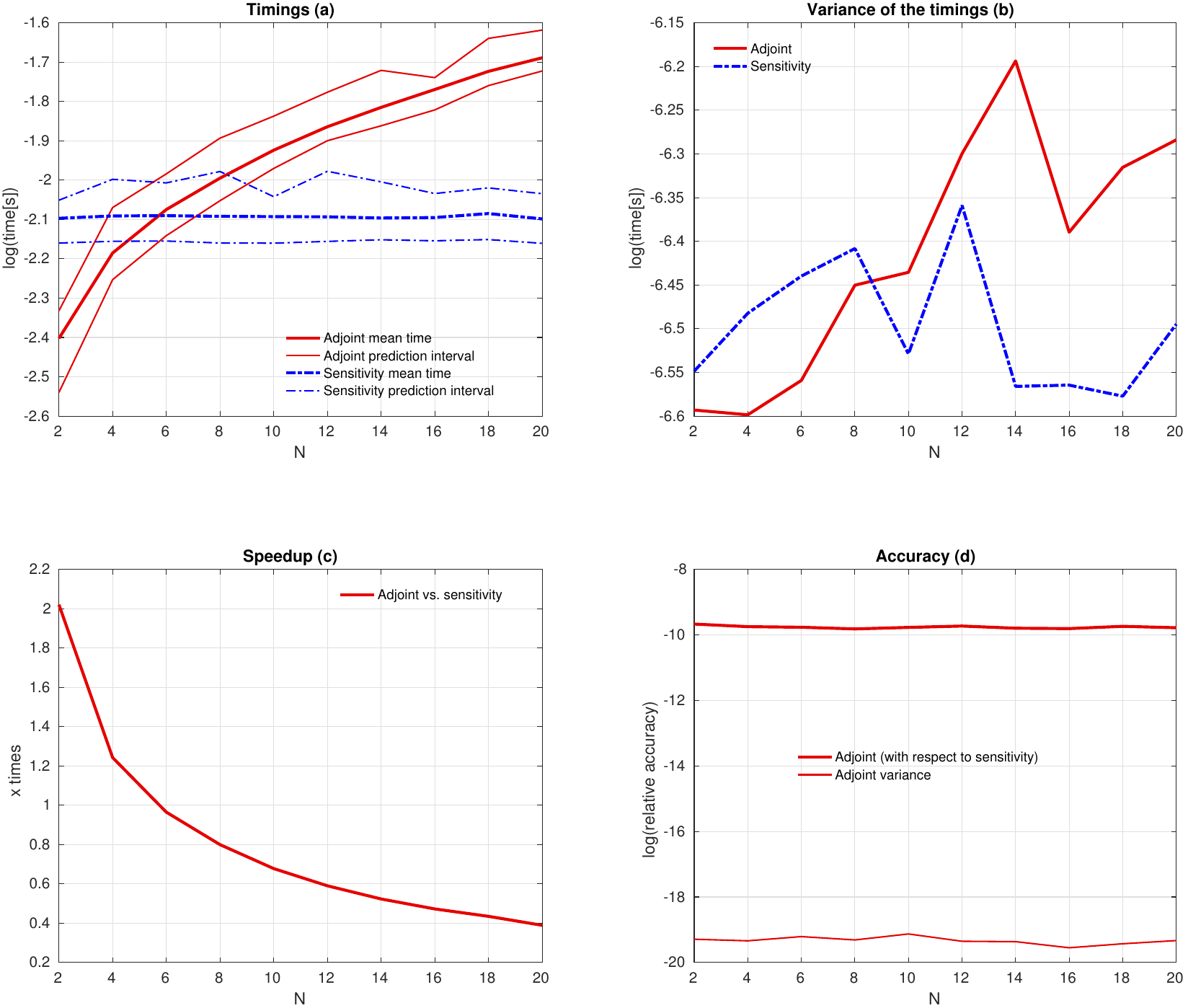}
\caption{HIV model: comparison of different methods for log-likelihood gradient computation with respect to the growing number of time observations}
\label{fig:gradientHIVnoto}
\end{figure}

For the HIV model $\dimpars$ is fixed and we can supply the results only with respect to $\nom.$ We again observe in Figure \ref{fig:gradientHIVnoto}(a) that the efficiency of \refe{eq:gradAdjoint} is strongly dependent on the number of observations. For up to 5 observations it is more efficient than the sensitivity equation (SE) approach. Thus even for models with a relatively small number of ODE parameters, the ASM approach for the computation of the gradient of \refe{eq:generallikelihood} can be advised for certain applications, such as mixed effects modeling in pharmacokinetics and pharmacodynamics, as in \citep{Lavielle2011}. However, for models with a few parameters and a high number of observation points, the SE approach is clearly more efficient. Accuracy-wise, both approaches are equivalent (Figure \ref{fig:gradientHIVnoto}(d)).

\begin{figure}
\includegraphics[width=\multfig\textwidth]{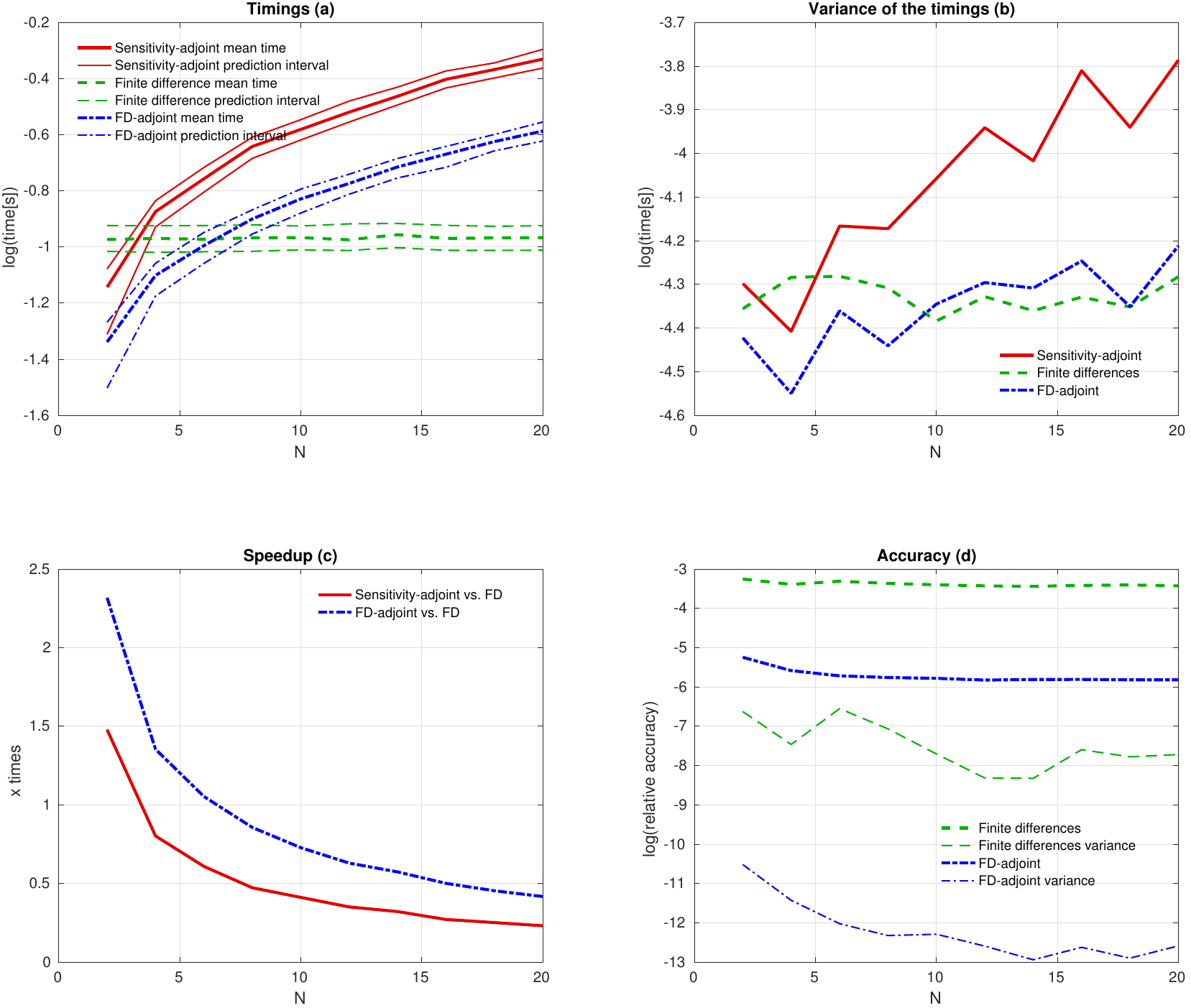}
\caption{HIV model: comparison of different methods for log-likelihood Hessian computation with respect to the growing number of time observations}
\label{fig:HessianHIVnoto}
\end{figure}

In Figure \ref{fig:HessianHIVnoto} the corresponding results for the Hessian computation of \refe{eq:generallikelihood} are presented. Three ways are compared: finite 
difference (FD) approximation, the ASM approach (SA) using \refe{eq:rsecder} and \refe{eq:SecDerSens} and the mixed approach (FA) using \refe{eq:FDadjoint}.

First, again as for the diagonal linear model in Section \ref{sec:linearModel}, the efficiency of the FD approximation is virtually independent of the number of measurements (Figure \ref{fig:HessianHIVnoto}(a)). This is not the case for the SA and FA approaches. However, due to its simplicity, the mixed FA approach is clearly more efficient than SA. It is more efficient than the FD approach up to 5 measurements, which corresponds to the previous results for the computation of the gradient.

The accuracies in Figure  \ref{fig:HessianHIVnoto}(d) are compared to the results of SA approach, as no exact solution is available. For the linear model \refe{eq:linear_model}, this approach was shown to be accurate up to the machine precision. The mixed FA approach achieves stable accuracies around $10^{-6},$ two orders of magnitude better then the full finite difference approximation.

\section{Conclusions}

We have derived and analyzed the adjoint-state method for computation of the gradient and the Hessian of likelihood functionals for time series data modeled by ordinary differential equations. We interfaced the discrete data and the continuous model on the level of likelihood functional, using the concept of point-wise distributions. The resulting adjoint problem \refe{eq:adjoint} then contains a Dirac delta source corresponding to individual measurement times. The developments are fully supported by the corresponding theoretical results. The implementation of a solver to \refe{eq:adjoint} closely follows the constructive proof of its well-posedness. 

Then, we compared the efficiency of the resulting ASM with finite differences and 
sensitivity equation (SE) approaches, both for the gradient and the Hessian. First, the implementation of SE approach is so efficient, that it renders the finite difference approximation practically obsolete, due to its superior accuracy. Second, the ASM efficiency is dependent on the number of measurement times, which is not the case for SE approach. For models with a high-number of parameters and a small number of measurement times, the ASM is a clear winner. It starts to be competitive even for rather small models like the latent dynamic HIV model from Section \ref{section:HIV} (11 parameters, 6 measurement times).

In future, we plan a sparse matrix code rewrite, which would allow for solution to bigger ODE systems and also a computationally more efficient implementation. Then, the preconditioning of Newton solver step during the CVODES integration of \refe{eq:adjoint} is an interesting possibility to speed up the ASM \citep{Knoll2004}.

\begin{acknowledgements}
We would like to thank Xavier Woot de Trixhe from Janssen Pharmaceutica for numerous very interesting discussions on PK/PD, virology, biological pathways modeling, NLMEMs and on life in general. They were an important source of motivation and provided a view from a different perspective.
\end{acknowledgements}

\section{Supplemental materials}
Two external files are provided:
\begin{enumerate}
\item A document titled: ``Supplemental material A: help with reproducing
of the results presented in Fast derivatives of
likelihood functionals for ODE based models using
adjoint-state method''.
\item A simple and extensively commented R-imple\-mentation of the ASM. It computes the gradient of the Gaussian log-likelihood \refe{eq:llfGaussian} with respect to the parameters of a pharmacokinetic two-compartment model. We set $\Omega=I$. The gradient is computed by the sensitivity equation method as well and those can be compared. But not in the terms of time efficiency. It is not a simple task to efficiently implement the ASM since the adjoint equation \refe{eq:adjoint} is integrated backward in time and the solver has to compute its r.h.s. dependent on the forward solution. This has to be cached by the solver. The R-solution uses a simple linear scheme which is far from optimal but illustrative. As described in Section \ref{sec:ASMimplementation}, the {\diffmem} implementation uses heavily the capabilities of the SUNDIALS package.

The R-code is independently understandable but nevertheless references the relevant formulas of this paper.
\end{enumerate}

\appendix
\section{Proofs}
\label{app:proofs}
\subsection*{Proof of Theorem \ref{theo:mainTheoremODE}}
First, when compared with Theorem 4.D from \cite{Zeidler1985} we work with $X=\RR^\dimmv$ and $P=\RR^\dimpars.$ Those are complete normed vector spaces i.e. they are Banach spaces. Second, the initial condition is dependent only on parameter $\pars$, not on any free parameter $y$ as in Theorem 4.D.

Set $J = [-1, 1].$ Let us do the following rescaling: $t = sa,$ $\bz(s) := \mv(as) - \mv_0(\pars)$ for all $s \in J.$ Then \refe{eq:ivp} is equivalent to 
\begin{equation*}
\bz'(s) - a\mrhs(as, \bz(s) + \mv_0(\pars), \pars) = \boldsymbol{0}\quad\mbox{for all } s \in J, \bz(0) = \boldsymbol{0}.
\end{equation*}
This can be written as an operator equation $F(\bz, a, \pars) = 0$ with the operator $F:\bZ\times \bA \to \bW$ and spaces $\bZ = \{\bz\in \bC^1(J,\RR^\dimmv): \bz(0) = \boldsymbol{0}\}$, $\bW = C(J,\RR^\dimmv).$ The space $\bA$ contains all the parameters $(a, \pars)$, i.e. $\bA = \RR \times \RR^\dimpars.$

Set $\bq = (\boldsymbol{0}, 0, \pars).$ Both $F$ and $F_z$ are continuous at $\bq.$ Obviously, $F(\bq)=\boldsymbol{0}$ and $F_{\bz}(\bq)\bz = \bz'.$  The crucial observation is that for every $\bw\in\bW,$ there exists exactly one $\bz\in\bZ$ with $\bz'=\bw$, namely $\bz(s)=\int_0^s \bw(t) \dt{t}$. Hence $F_{\bz}(\bq):\bZ \to \bW$ is bijective. The implicit function theorem yields the conclusions (see e.g. Theorem 4.B in \cite{Zeidler1985}).
$\blacksquare$

\subsection*{Proof of Lemma \ref{lemma:gradSensitivity}}
After realizing that $\llf$ depends on $\pars$ only through $\mv$, \refe{eq:rfirstder} is formally a direct application of the chain rule ($\dist_{\mv}$ denotes the derivative of the metric with respect to the model state $\mv.$)

The r.h.s. of \refe{eq:rfirstder} is a well-posed finite expression. First, we have assumed that the metric is sufficiently smooth, thus $\dist_{\mv}$ is continuous. Second, $\sv$ is continuous as well owing to Theorem \ref{theo:mainTheoremODE}. A distribution can be rescaled by any at least continuous function, as here $\{\delta\}$ by $\dist_{\mv}$.

Thus, the first differential on the l.h.s of \refe{eq:rfirstder} exists as well. It is moreover continuous, i.e. it is Fr\'echet. \footnote{An alternative argumentation could employ equivalence \refe{eq:diraccost}.} $\blacksquare$

\subsection*{Proof of Theorem \ref{theo:gradAdjoint}}
First, let us assume that we have already constructed a unique solution $\dv$ to \refe{eq:adjoint} up to a certain measurement point $t_i$. The adjoint problem is solved backwards in time. Consequently, we will construct its prolongation on $[t_i, t_{i-1})$.

Let $\dv_i^+$ be the ODE solution just before integrating 
the measurement at time $t_i$, i.e. at time $t_i^+.$ We simply stop 
the integration at $t_i^+,$ add $\dist_{\mv}(\meas_i,\mf(t_i, \pars))$ to $\dv_i^+$ and solve 
\begin{equation}
\begin{aligned}
\der{\dv}{t} &= -\jact{\mrhs}{\mv}\dv, 
\quad t\in (t_i,t_{i-1}), \\ 
\dv(t_{i}) &= \dv_i^+ + \dist_{\mv}(\meas_i,\mf(t_i, \pars)).
\end{aligned}
\label{eq:adjointWithoutDirac}
\end{equation}
This is a simple linear ODE with a continuous coefficient $\jact{\mrhs}{\mv},$ since $f\in \bC^1$. The classical results yield the global solution on $(t_i,t_{i-1})$ (see e.g. Theorem 5.1 and Theorem 5.2 from \citep{Coddington1955}.) This concludes the proof of the existence and uniqueness.

Now, we prove \refe{eq:rfirstderAdjoint}. Let us without a loss of generality assume that there are no measurements in times $0$ and $T.$ It is a well-known result of theory of distributions (in the sense of functional analysis), that the classical integration by part formula 
\begin{equation}
\int_0^T \der{\dv}{t} \bw \ dt = [\dv\bw]_{0}^{T} -\int_0^T \dv \der{\bw}{t} \ 
dt
\label{eq:bypartsdelta}
\end{equation}
is valid for $\bw\in \bC^1$ even if the derivative $\der{\dv}{t}$ exists on $[0,T]$ only in a \emph{weak} sense, i.e. almost everywhere. Actually, \refe{eq:bypartsdelta} is the definition of the weak derivative of $\dv$ taking only $\bw \in \bC^1_0([0,T]).$ Consequently, since $\sv\in \bC^1([0,T]),$ we can safely proceed as follows
\begin{equation}
\begin{aligned}
\dual{\sv}{\{\delta\}\dist_{\mv}(\meas,\mf(t, 
\pars))}
&\mathop{=}\limits^{\mbox{\scriptsize\refe{eq:adjoint}}}
\dual{\sv}{\der{\dv}{t}
+ \jact{\mrhs}{\mv}\dv}\\
&\mathop{=}\limits^{\mbox{\scriptsize\refe{eq:sensitivity}}}  -\dv^t(0)\jac{\mv_0}{\pars}\bh -\scal{\der{\sv}{t} - 
\jac{\mrhs}{\mv}\sv}{\dv} \\
&\mathop{=}\limits^{\mbox{\scriptsize\refe{eq:sensitivity}}} 
-\dv^t(0)\jac{\mv_0}{\pars}\bh-\scal{\jac{\mrhs}{\pars}\var{
\pars}}{\dv}.\blacksquare
\end{aligned}
\end{equation}

\subsection*{Proof of Lemma \ref{lemma:HessianAdjoint}}
The existence and uniqueness of $\ssv$ is a direct results of Theorem \ref{theo:mainTheoremODE}. Now, \refe{eq:SecDerSens} is derived as follows:
\begin{equation}
\begin{aligned}
\dual{\ssv}{\{\delta\}\dist_{\mv}(\meas,\mf(t, \pars))}
&\mathop{=}\limits^{\refe{eq:adjoint}}
\dual{\ssv}{\der{\dv}{t} + \jact{\mrhs}{\mv}\dv}\\
&\mathop{=}\limits^{\refe{eq:adjoint},\refe{eq:secondsens}}
-\pddl{(\mv_0)}{\pars\pars}\bh_1\bh_2\cdot\dv(0)\\
&-\scal{\der{\ssv}{t}}{\dv}+\scal{\jac{\mrhs}{\mv}\ssv}{
\dv}.
\end{aligned}
\end{equation}
This after substituting for $\jac{\mrhs}{\mv}\ssv$ from \refe{eq:secondsens} directly yields \refe{eq:SecDerSens}. Analogically to the proof of Theorem \ref{theo:gradAdjoint}, we needed $\ssv \in \bC^1([0,T])$ to be able to integrate by parts.$\blacksquare$

\end{document}